\documentclass[prd,twocolumn,showpacs,preprintnumbers,amsmath,amssymb,superscriptaddress,nofootinbib,english]{revtex4-1}

\usepackage{graphicx}% Include figure files
\usepackage{dcolumn}% Align table columns on decimal point
\usepackage{bm}% bold math
\usepackage{epsfig}
\usepackage{graphicx}
\usepackage{hyperref}
\usepackage[usenames]{color}
\usepackage{url}
\usepackage{enumitem}

\hypersetup{
    colorlinks=true,
    linkcolor=red,
    citecolor=blue,
}

\newcommand{\tc}{\textcolor{black}}

\newcommand{\remove}[1]{}

\def\etal{{\frenchspacing\it et al.}}

\def\eg{{\frenchspacing\it e.g.}}

\def\be{\begin{equation}}
\def\ee{\end{equation}}
\def\ba{\begin{eqnarray}}
\def\ea{\end{eqnarray}}
\begin{document}
\title{Probing dynamics of dark energy with latest observations}

\author{Yuecheng Zhang}
\affiliation{National Astronomy Observatories,
Chinese Academy of Science, Beijing, 100012, P.R.China}
\affiliation{University of Chinese Academy of Sciences, 
Beijing, 100049, P.R.China}

\author{Hanyu Zhang}
\affiliation{National Astronomy Observatories,
Chinese Academy of Science, Beijing, 100012, P.R.China}

\author{Dandan Wang}
\affiliation{National Astronomy Observatories,
Chinese Academy of Science, Beijing, 100012, P.R.China}
\affiliation{University of Chinese Academy of Sciences, 
Beijing, 100049, P.R.China}

\author{Yanghan Qi}
\affiliation{Department of Physics \& Astronomy, Swarthmore College, 
Swarthmore, PA 19081 USA}

\author{Yuting Wang}
\affiliation{National Astronomy Observatories,
Chinese Academy of Science, Beijing, 100012, P.R.China}
\affiliation{Institute of Cosmology and Gravitation, University of Portsmouth,
Portsmouth, PO1 3FX, UK}

\author{Gong-Bo Zhao}
\email{gbzhao@nao.cas.cn}
\affiliation{National Astronomy Observatories,
Chinese Academy of Science, Beijing, 100012, P.R.China}
\affiliation{University of Chinese Academy of Sciences, 
Beijing, 100049, P.R.China}
\affiliation{Institute of Cosmology and Gravitation, University of Portsmouth,
Portsmouth, PO1 3FX, UK}

\begin{abstract}

We examine the validity of the $\Lambda$CDM model, and probe for the dynamics of dark energy using latest astronomical observations. Using the $Om(z)$ diagnosis, we find that different kinds of observational data are in tension within the $\Lambda$CDM framework. We then allow for dynamics of dark energy and investigate the constraint on dark energy parameters. We find that for two different kinds of parametrisations of the equation of state parameter $w$, a combination of current data mildly favours an evolving $w$, although the significance is not sufficient for it to be supported by the Bayesian evidence. A forecast of the DESI survey shows that the dynamics of dark energy could be detected at $7\sigma$ confidence level, and will be decisively supported by the Bayesian evidence, if the best fit model of $w$ derived from current data is the true model. 

\end{abstract}


\maketitle

\section{Introduction}

The accelerating expansion of the Universe revealed by supernovae type Ia (SNIa) is one of the most significant discoveries in modern cosmology \cite{RiessPerl}. In the framework of general relativity, the cosmic acceleration in the late Universe is due to dark energy (DE), a yet unknown energy component contributing to about two thirds of the total energy budget of the Universe. From astronomical observations, measurements of the equation of state parameter (EoS) $w$, which is the ratio of pressure to energy density of DE, can shed light on the nature of DE as different DE models can be characterised by $w$. For example, the cosmological constant $\Lambda$, which is one of the most popular DE models, predicts that $w=-1$, while in dynamical dark energy (DDE) models including {\it quintessence} \cite{quintessence}, {\it phantom} \cite{phantom}, {\it quintom} \cite{quintom} and so on, $w$ evolves with redshift $z$. Hence reconstructing the $w(z)$ function from observations including cosmic microwave background (CMB), SNIa and large scale structure (LSS) measurements, is an efficient way to test dark energy models.

Performing a consistency check for the $\Lambda$CDM model, which has least number of model parameters compared with DDE models in general, using observations is a common starting point for phenomenological studies of dark energy. Interestingly, recent studies show that different kinds of observational data are in tension within the framework of the $\Lambda$CDM model \cite{Zhao17,Font-Ribera:2013wce,Sahni:2014ooa,Battye:2014qga,Aubourg:2014yra,P15,Addison:2015wyg,Bernal:2016gxb,Sola1,Sola2,DMS, DMLS}. In particular, Zhao \etal~(2017) \cite{Zhao17} quantifies the tension using the Kullback-Leibler divergence \cite{Kullback:1951va}, and uses a nonparametric DDE model to successfully relieve the tension. Their analysis basically shows that the tension within $\Lambda$CDM can be interpreted as a signal of dynamics of dark energy at a $3.5\sigma$ confidence level (CL). 

In this paper, we perform a complementary study to Zhao \etal~(2017). We first reinvestigate the tension between different datasets using the $Om$ \cite{Om1,Om2} diagnosis, and then reconstruct $w(z)$ following a parametric approach. We quantify the significance of $w\ne-1$ and perform a model selection using the Bayesian evidence on current and simulated future observational data.

This paper is organised as follows. In the next section we present the method and datesets used, and in section III we present the result, followed by a section of conclusion and discussion.

\begin{table*}[htbp]  
\begin{center}
\begin{tabular}{ccccc|cc}
\hline\hline
			\multicolumn{7}{c}{Parametrisation I}									\\
\hline
                	$w_0$		&$w_1$		&	$w_2$		&$w_3$ 	&	$w_4$ & $\sqrt{|\Delta\chi^2|}$ & $\Delta{\rm ln}E$ 	\\					
                 \hline

		$-1.02\pm0.04(0.01)$		&$0$				&$0$				&$0$ 			&$0$				&$0.4(0.8)$		&$-2.3(-3.4)\pm0.3$  	\\	
		$-1.08\pm0.10(0.05)$		&$0.26\pm0.40(0.21)$	&$0$				&$0$ 			&$0$				&$0.7(5.2)$ 		&$-3.9(6.1)\pm0.3$   \\
		$-1.18\pm0.17(0.08)$		&$1.50\pm1.75(0.67)$	&$-2.34\pm3.21(1.14)$	&$0$ 			&$0$				&$1.1(5.4)$ 			&$-7.1(3.1)\pm0.3$   \\
		$-1.07\pm0.17(0.10)$		&$-1.42\pm2.40(1.22)$	&$12.1\pm10.2(3.75)$	&$-17.7\pm12.6(3.32)$	&$0$				&$1.8(5.6)$			&$-8.4(0.5)\pm0.3$  \\
		$-1.00\pm0.18(0.09)$		&$0.38\pm2.72(1.59)$	&$-15.8\pm21.2(9.29)$	&$72.0\pm62.3(20.0)$ 	&$-79.6\pm55.0(13.4)$	&$2.2(6.0)$ 			&$-8.8(0.0)\pm0.3$  \\
\hline\hline

		 \multicolumn{7}{c}	{Parametrisation II} 		\\
\hline
		$w_0$		&$w_1$		&	$w_2$		&$w_3$ &	$w_4$ & $\sqrt{|\Delta\chi^2|}$ & $\Delta{\rm ln}E$ \\
                 \hline
		$-1.03\pm0.04(0.03)$		&$4.98\pm2.87(0.61)$		&	$5.38\pm2.43(0.39)$		&$13.3\pm6.42(0.40)$ 	&$0$ 			&$2.6(7.4)$ 	&$-2.2(14.0)\pm0.3$\\
		$-1.03\pm0.05(0.03)$		&$4.77\pm2.86(0.64)$		&	$5.61\pm2.46(0.41)$		&$13.8\pm7.57(0.84)$ 	&$4.90\pm2.84(1.82)$ 	&$2.6(7.5)$  	&$-2.0(14.2)\pm0.3$\\
\hline\hline
\end{tabular}
\end{center}
\label{tab:table}
\caption{Constraints on dark energy parameters using current data and simulated data (numbers quoted in parenthesis) respectively.}
\end{table*}%

\section{Method and Data}

In this section, we present the methodology used for quantifying the tension among datasets, for performing dark energy model parameter inference and for model selection. We also describe datasets used in this work.

\subsection{The $Om$ diagnosis}

The quantity $Om$ is defined as follows \cite{Om1,Om2},
\be 
Om(z) \equiv \frac{[H(z)/H_0]^2-1}{(1+z)^3-1}.
\ee
where $H(z)$ and $H_0$ are the Hubble parameter measured at redshift $z$ and $0$ respectively. It is a useful diagnosis of any deviation from the $\Lambda$CDM model simply because $Om(z)=\Omega_m$ in $\Lambda$CDM. Thus any non-constancy of $Om(z)$ signals that $w\ne-1$, if the flatness of the Universe is assumed.

Observationally, $H_0$ can be directly measured in the local Universe, and $H(z)$ can be estimated from CMB, baryonic acoustic oscillations (BAO) redshift surveys using either galaxies (gBAO), or Lyman-$\alpha$ forest (Ly$\alpha$FB), or from the relative age of old and passively evolving galaxies following a cosmic chronometer approach (OHD).
 
 \begin{figure}[tbp]   % * for total page%
\includegraphics[scale=0.3]{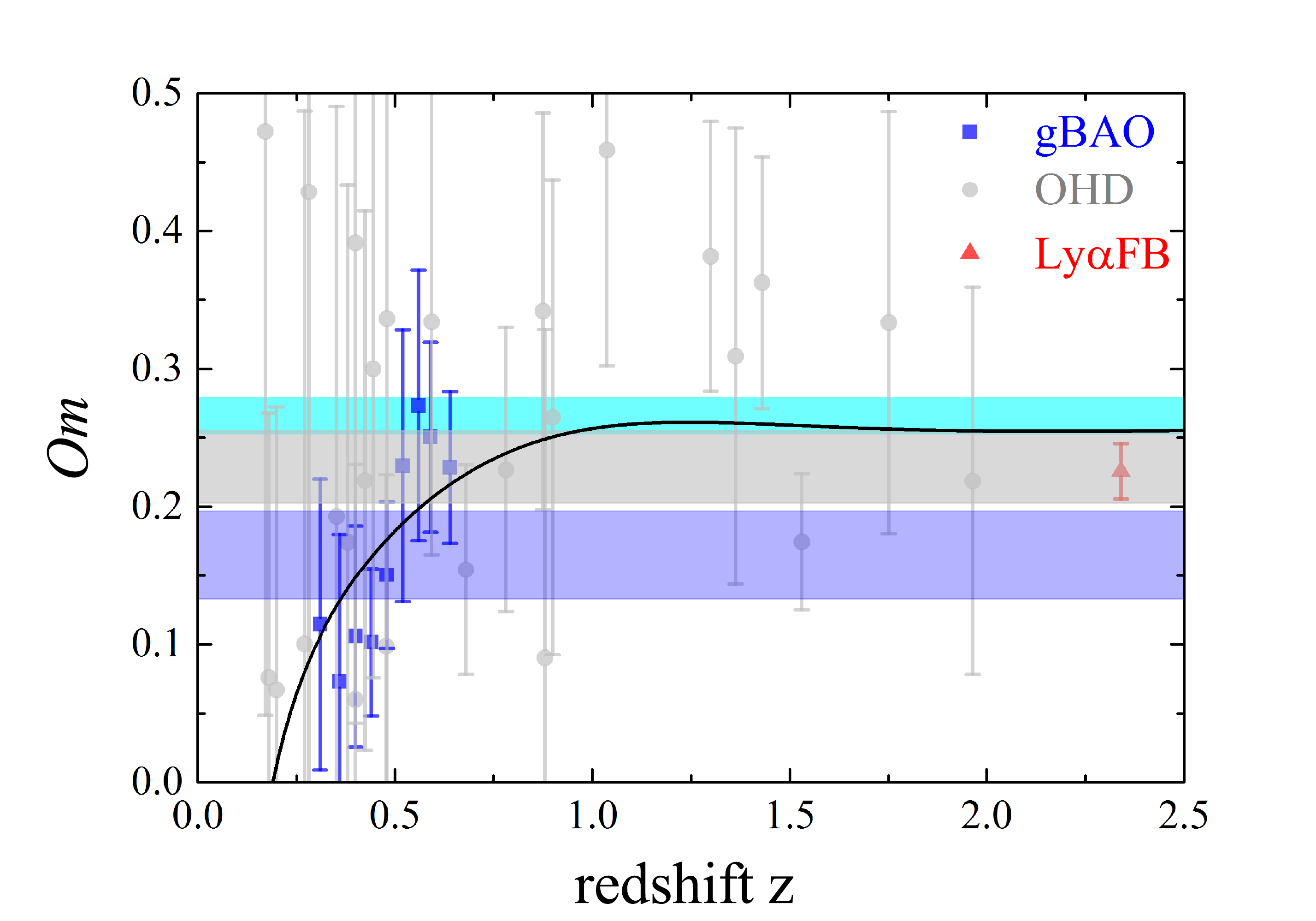}
\caption{The measured $Om$ from various kinds of data: galaxy BAO (blue square), OHD (grey circle) and Lyman-$\alpha$ forest BAO (red triangle). The horizontal cyan, grey and blue bands show the 68\% CL allowed values for a constant $Om$ fitted to Planck 2015, OHD and Ly$\alpha$FB respectively. The black solid curve shows $Om$ derived from the best fit $w(z)$ model. See texts for details.}\label{fig:om}
\end{figure}

\begin{figure*}[tbp]   % * for total page%
\includegraphics[scale=0.27]{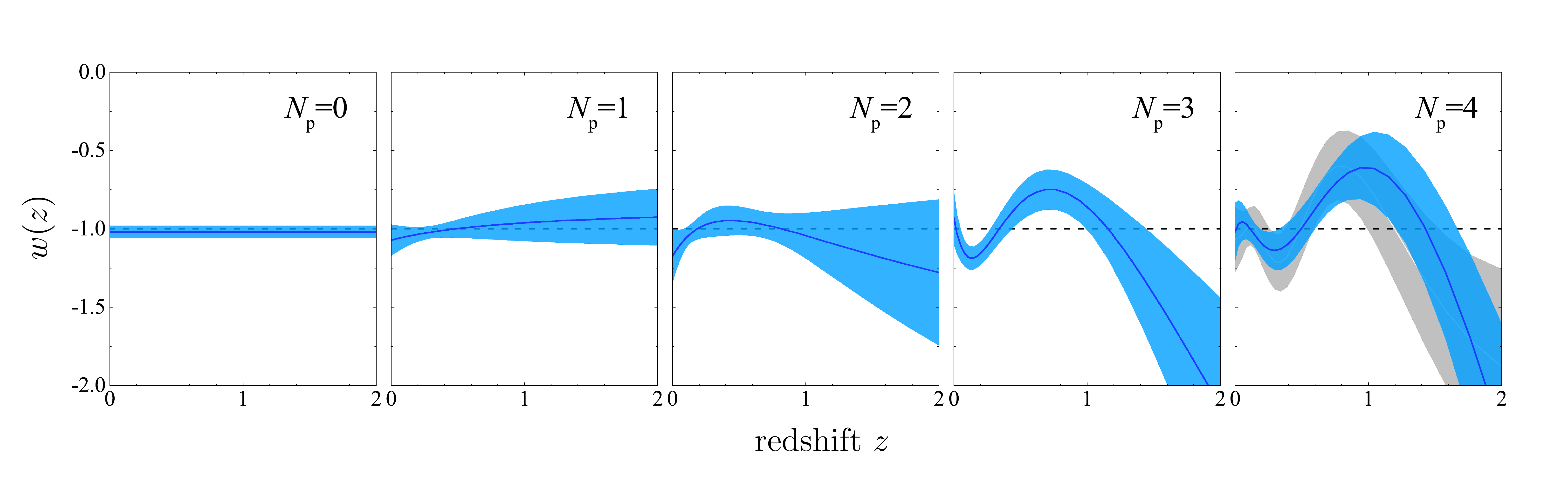}
\caption{Blue bands: the mean with 68\% CL error of the reconstructed $w(z)$ using parametrisation I for different orders of the polynomial. The grey band in the $N_{\rm p}=4$ panel shows the nonparametric $w(z)$ reconstruction result in Zhao et al.~(2017).}\label{fig:poly}
\end{figure*}

\begin{figure}[tbp]   % * for total page%
\includegraphics[scale=0.27]{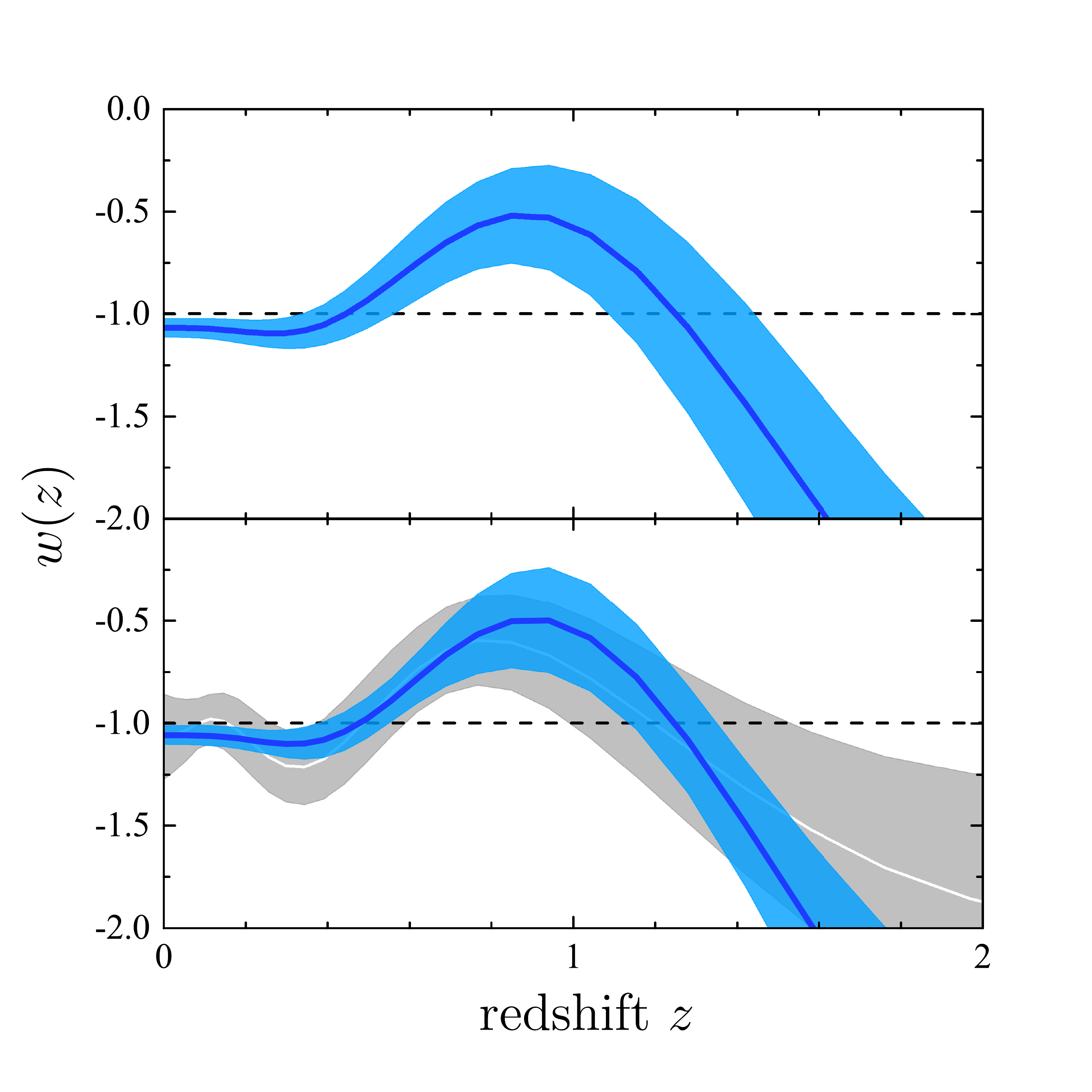}
\caption{Same as Fig. \ref{fig:poly} but for parametrisation II. The upper and lower panels show the reconstruction result with and without the $w_4$ parameter fixed respectively.}\label{fig:osc}
\end{figure} 
 
\subsection{Parametrisations of the Universe} 

In this work, we consider two kinds of parametrisations of $w(a)$, where $a$ is the scale factor of the Universe\footnote{For more parametrisations of $w(a)$, see \cite{PNP}.}. 

\begin{description}[leftmargin=\parindent, labelwidth=\parindent, labelsep=0pt, align=left]
\item [Parametrisation I]~~Polynomial expansion \cite{Ade:2015rim}
\be\label{eq:poly}
w(a) = \sum_{i=0}^{N_{\rm p}} w_i(1-a)^{i} 
\ee
where $N_{\rm p}$ defines the order of the polynomial expansion. Note that $N_{\rm p}=0$ and $N_{\rm p}=1$ are the $w$CDM model, in which $w$ is a constant, and the Chevallier-Polarski-Linder (CPL) model \cite{CP,L} respectively, and including higher order terms allows more general behaviour of $w(a)$. In this work, we consider cases with $N_{\rm p}\leqslant4$.

\item [Parametrisation II]~~Oscillatory function

Although Parametrisation I allows for oscillatory behaviours of $w(a)$ in general, it requires a large number of terms in order to properly approximate a periodic oscillatory function, \eg, a cosine function. Therefore we consider another kind of parametrisation as,
\be\label{eq:osc}
w(a) = w_0 + w_1(1-a)^{w_2} {\rm cos} \left(w_3 a + w_4\right)    
\ee
This is a general cosine function that allows its mean, amplitude, period and phase to be free parameters. It is similar to the functional form used in \cite{Feng:2004ff} but is more general in that the $(1-a)^{w_2}$ term allows the amplitude to vary with the scale factor.
\end{description}

Our parametrisation of the Universe is thus, \be\label{eq:para} {\bf P}\equiv\{ \omega_{b}, \omega_{c}, \Theta_{s}, \tau, n_s, A_s, w_0, ..., w_{4},\mathcal{N}) \ee  where $\omega_{b}$ and $\omega_{c}$ are the baryon and cold dark matter physical densities, $\Theta_{s}$ is the angular size of the sound horizon at decoupling, $\tau$ is the optical depth, $n_s$ and $A_s$ are the spectral index and the amplitude of the primordial power spectrum, and $w_0,...,w_{4}$ denote the above-mentioned dark energy EoS parameters. We marginalize over nuisance parameters $\mathcal{N}$ such as the intrinsic SN luminosity, galaxy bias, etc.
 
\subsection{Observational datasets used} 
\label{sec:data}
 
The datasets we consider in this work include the gBAO measurements that utilize the BOSS DR12 sample at nine effective redshifts \cite{BAOZhao, BAOWang}, the Ly$\alpha$FB measurements \cite{Delubac14}, the 6dFRS \cite{6df} and SDSS main galaxy sample \cite{MGS} BAO measurements, the WiggleZ galaxy power spectra \cite{wigglez},  the recent estimate of the Hubble constant $H_0$ obtained from local measurements of Cepheids \cite{R16} ($H_0$), the recent OHD measurements of $H(z)$ \cite{hz16}, the JLA sample of SNIa \cite{Betoule:2014frx},  the weak lensing shear angular power spectra from CFHTLenS \cite{WL} and the {\it Planck} 2015 CMB temperature and polarisation angular power spectra \cite{P15}.

For the purpose of forecast, we simulate future gBAO data assuming a DESI \footnote {\url{http://desi.lbl.gov/}} sensitivity following \cite{DESI}, and also consider a future space-based supernova mission described in \cite{EuclidSN}. 

\subsection{Parameter estimation and model selection}

We use a modified version of CAMB \cite{CAMB} to calculate observables, and include dark energy perturbations following the approach developed in \cite{DEP}. We perform a Markov Chain Monte Carlo (MCMC) global fitting of parameters listed in Eq (\ref{eq:para}) to a combination of datasets described in Sec. \ref{sec:data} using a modified version of CosmoMC \cite{Lewis:2002ah}, and use the PolyChord \cite{polychord} plug-in of CosmoMC to compute the Bayesian evidence for the model selection. 

\section{Result}

We present our results in Table \ref{tab:table} and in Figs \ref{fig:om}-\ref{fig:osc}.

The quantity $Om(z)$ is estimated using $H(z)$ measurements from {\it Planck} 2015, gBAO, OHD and Ly$\alpha$FB respectively, with the recent $H_0$ measurement presented in \cite{R16}. To check the constancy of $Om(z)$ using each individual kind of datasets, and the consistency between different kinds of data, we fit constants to the $Om(z)$ measurements from {\it Planck} 2015, gBAO and OHD separately, and show the 68\% CL constraints in cyan, blue and grey horizontal bands respectively in Fig \ref{fig:om}. Specifically, we obtain,
\ba Om({\rm {\it Planck}~2015})&=& 0.266\pm0.013 \\
      Om({\rm gBAO})&=& 0.165\pm0.032 \\
      Om({\rm OHD}) &=& 0.229\pm0.026 \\ 
      Om({\rm Ly\alpha FB}) &=& 0.226\pm0.020\ea
It is true that neither the {\it Planck} 2015, gBAO nor OHD dataset shows a significant deviation from a constant $Om$ given the level of uncertainty, however, the derived $Om$'s from {\it Planck} 2015, gBAO and OHD are different at larger than $2\sigma$ CL. Furthermore, the $Om$ values derived here are all smaller than $\Omega_m$ derived from {\it Planck} 2015 alone in the $\Lambda$CDM model \cite{P15}, which is $\Omega_m=0.315\pm0.013$. This to some extent is due to the fact that the $H_0$ value used here, which is $73.24\pm1.74$ km~${\rm s}^{-1}{\rm Mpc}^{-1}$, is significantly larger than that derived from {\it Planck} 2015, which is $67.31\pm0.96$ km~${\rm s}^{-1}{\rm Mpc}^{-1}$. All these discrepancy among datasets suggests that the $\Lambda$CDM model may need to be extended.

For more general DE models parametrised by Eqs (\ref{eq:poly}) and (\ref{eq:osc}), we derive constraints on model parameters, which are shown in Table \ref{tab:table}. For the polynomial expansion case, we increasingly add higher order terms to the $w$CDM model in the global fitting. We find that the $\chi^2$ can be reduced by $4.8$ at most for the $N_{\rm p}=4$ model. For the purpose of model selection, we also evaluate the logarithmic Bayesian factor, \be\Delta{\rm ln}E \equiv {\rm ln}E_{{\rm DDE}}- {\rm ln}E_{\Lambda{\rm CDM}}\ee where \be E\equiv \int d^{n}\theta P(\theta)\ee denotes the Bayesian evidence, which is an integral of the probability distribution function of $n$-dimensional parameters $\theta$. We find that $\Delta{\rm ln}E$ is negative for all cases, meaning that neither of these DDE models is favoured over the $\Lambda$CDM model. For the $N_{\rm p}=4$ case, in which $w(z)$ is parametrised with five free parameters, is found to be not equal to $-1$ at $2.2\sigma$ CL, and the Bayesian factor is as low as $\Delta{\rm ln}E=-8.8\pm0.3$, which strongly indicates current data do not support extending $\Lambda$ in this parametrisation.

For parametrisation II, we show results with and without the phase $w_4$ fixed, and find that whether $w_4$ varies or not does not change the result: $\chi^2$ is reduced by $6.8$ (a $2.6\sigma$ signal of $w\ne-1$) by four additional parameters with a Bayesian factor $\Delta{\rm ln}E=-2.2\pm0.3$. Admittedly, although this model is also not supported by the Bayesian evidence, it is much less disfavoured than the $N_{\rm p}=4$ model in parametrisation I, and it fits to the data better. 

In Figs \ref{fig:poly} and \ref{fig:osc}, we reconstruct $w(z)$ using constraints on DE parameters we obtained. As shown, the best fit $w(z)$ models with all five DE parameters varied, which are shown in the far right panel of Fig \ref{fig:poly}, and in the lower panel in Fig \ref{fig:osc}, crosses $-1$ during evolution, and exhibits certain level of oscillations with respect to redshift $z$, which is consistent with the prediction of the model of {\it oscillating quintom} \cite{Feng:2004ff}. We compare this result to the nonparametric reconstruction presented in \cite{Zhao17}. As shown, our result is consistent with that in Zhao \etal~(2017) within $1\sigma$ CL. 

To reinvestigate the tension among various datasets in DDE models, we over-plot $Om$ for the best fit DDE model as parametrised by Eq (\ref{eq:osc}) (black solid). As shown, it is consistent with all datasets, signalling a release of tension among datasets.

To assess whether the best fit $w$ model found in this work will be supported by future observations, we take the best fit $w$ model as a fiducial model, create mock BAO and supernovae data assuming a DESI \cite{DESI} and a future space-based supernova mission \cite{EuclidSN} combined with {\it Planck} 2015 data, and repeat our analysis. We find that for parametrisation I, models of $N_{\rm p}=1,2$ will be supported by Bayesian evidences, with a signal of $w\ne-1$ at $5\sigma$ CL. Although the $N_{\rm p}=3, 4$ models fit data better, they are not much preferred to the $\Lambda$CDM model even for the future data. On the other hand, future data support the oscillation model much more significantly. Namely, those models will be detected at more than $7\sigma$ CL with a large Bayesian factor of $\Delta{\rm ln}E=14\pm0.3$.
 
\section{Conclusion and Discussions}

We revisit the consistency among various kinds of recent observations using the $Om$ diagnosis, and confirm that the tension exists among {\it Planck} 2015, gBAO, OHD, Ly$\alpha$FB and the new $H_0$ measurement in the $\Lambda$CDM model.

We therefore allow the dynamics of dark energy and perform parametric reconstruction of $w(z)$ with two kinds of parametrisations using a combination of current datasets, and using the simulated future data. We find that an oscillatory $w(z)$ across $-1$ during the evolution is mildly favoured by a combination of current observations at a confidence level of $2.6\sigma$ based on the improvement in $\chi^2$. This model can well relieve the tension among datasets. It is true that this is not sufficient for it to be supported by the Bayesian evidence, however, for future galaxy surveys with a sensitivity similar to DESI and space-based supernova surveys, the best-fit model derived in this work will be detected at a confidence level of $7\sigma$, and will be decisively supported by the Bayesian evidence.

\begin{acknowledgments}
The authors are supported by NSFC Grant No. 11673025, and by a Key International Collaboration Grant from Chinese Academy of Sciences. GBZ is also supported by a Royal Society Newton Advanced Fellowship.\end{acknowledgments}


\begin{thebibliography}{99}

\bibitem{RiessPerl} 

  A.~G.~Riess {\it et al.} [Supernova Search Team],
  %``Observational evidence from supernovae for an accelerating universe and a cosmological constant,''
  Astron.\ J.\  {\bf 116}, 1009 (1998)
  %doi:10.1086/300499
  [astro-ph/9805201];
    S.~Perlmutter {\it et al.} [Supernova Cosmology Project Collaboration],
  %``Measurements of Omega and Lambda from 42 high redshift supernovae,''
  Astrophys.\ J.\  {\bf 517}, 565 (1999)
 % doi:10.1086/307221
  [astro-ph/9812133].



\bibitem{quintessence} 
  B.~Ratra and P.~J.~E.~Peebles,
  %``Cosmological Consequences of a Rolling Homogeneous Scalar Field,''
  Phys.\ Rev.\ D {\bf 37}, 3406 (1988).
  doi:10.1103/PhysRevD.37.3406;
  %%CITATION = doi:10.1103/PhysRevD.37.3406;%%
  %3045 citations counted in INSPIRE as of 21 Mar 2017
  P.~J.~E.~Peebles and B.~Ratra,
  %``Cosmology with a Time Variable Cosmological Constant,''
  Astrophys.\ J.\  {\bf 325}, L17 (1988).
  doi:10.1086/185100
  %%CITATION = doi:10.1086/185100;%%
  %1357 citations counted in INSPIRE as of 21 Mar 2017



%\cite{Caldwell:1999ew}
\bibitem{phantom} 
  R.~R.~Caldwell,
  %``A Phantom menace?,''
  Phys.\ Lett.\ B {\bf 545}, 23 (2002)
  doi:10.1016/S0370-2693(02)02589-3
  [astro-ph/9908168].
  %%CITATION = doi:10.1016/S0370-2693(02)02589-3;%%
  %2096 citations counted in INSPIRE as of 21 Mar 2017

\bibitem{quintom}
  B.~Feng, X.~L.~Wang and X.~M.~Zhang,
  %``Dark Energy Constraints from the Cosmic Age and Supernova,''
  Phys.\ Lett.\  B {\bf 607}, 35 (2005)
  [arXiv:astro-ph/0404224].
  

\bibitem{Zhao17} 
  G.~B.~Zhao {\it et al.},
  %``The clustering of galaxies in the completed SDSS-III Baryon Oscillation Spectroscopic Survey: Examining the observational evidence for dynamical dark energy,''
  arXiv:1701.08165 [astro-ph.CO].
  %%CITATION = ARXIV:1701.08165;%%

\bibitem{Font-Ribera:2013wce} 
  A.~Font-Ribera {\it et al.} [BOSS Collaboration],
  %``Quasar-Lyman $\alpha$ Forest Cross-Correlation from BOSS DR11 : Baryon Acoustic Oscillations,''
  JCAP {\bf 1405}, 027 (2014)
%  doi:10.1088/1475-7516/2014/05/027
  [arXiv:1311.1767 [astro-ph.CO]].
  %%CITATION = doi:10.1088/1475-7516/2014/05/027;%%
  %72 citations counted in INSPIRE as of 10 Jul 2016  

%\cite{Sahni:2014ooa}
\bibitem{Sahni:2014ooa} 
  V.~Sahni, A.~Shafieloo and A.~A.~Starobinsky,
  %``Model independent evidence for dark energy evolution from Baryon Acoustic Oscillations,''
  Astrophys.\ J.\  {\bf 793}, no. 2, L40 (2014)
 % doi:10.1088/2041-8205/793/2/L40
  [arXiv:1406.2209 [astro-ph.CO]].
  %%CITATION = doi:10.1088/2041-8205/793/2/L40;%%
  %60 citations counted in INSPIRE as of 25 Jan 2017

%\cite{Battye:2014qga}
\bibitem{Battye:2014qga} 
  R.~A.~Battye, T.~Charnock and A.~Moss,
  %``Tension between the power spectrum of density perturbations measured on large and small scales,''
  Phys.\ Rev.\ D {\bf 91}, no. 10, 103508 (2015)
  %doi:10.1103/PhysRevD.91.103508
  [arXiv:1409.2769 [astro-ph.CO]].
  %%CITATION = doi:10.1103/PhysRevD.91.103508;%%
  %33 citations counted in INSPIRE as of 07 Jan 2017

%\cite{Aubourg:2014yra}
\bibitem{Aubourg:2014yra} 
  \'E.~Aubourg {\it et al.},
  %``Cosmological implications of baryon acoustic oscillation measurements,''
  Phys.\ Rev.\ D {\bf 92}, no. 12, 123516 (2015)
  %doi:10.1103/PhysRevD.92.123516
  [arXiv:1411.1074 [astro-ph.CO]].
  %%CITATION = doi:10.1103/PhysRevD.92.123516;%%
  %124 citations counted in INSPIRE as of 25 Jan 2017

%\cite{Ade:2015xua}
\bibitem{P15} 
  P.~A.~R.~Ade {\it et al.} [Planck Collaboration],
  %``Planck 2015 results. XIII. Cosmological parameters,''
  Astron.\ Astrophys.\  {\bf 594}, A13 (2016)
 % doi:10.1051/0004-6361/201525830
  [arXiv:1502.01589 [astro-ph.CO]].
  %%CITATION = doi:10.1051/0004-6361/201525830;%%
  %2550 citations counted in INSPIRE as of 10 Dec 2016
  
  %\cite{Addison:2015wyg}
\bibitem{Addison:2015wyg} 
  G.~E.~Addison, Y.~Huang, D.~J.~Watts, C.~L.~Bennett, M.~Halpern, G.~Hinshaw and J.~L.~Weiland,
  %``Quantifying discordance in the 2015 Planck CMB spectrum,''
  Astrophys.\ J.\  {\bf 818}, no. 2, 132 (2016)
%  doi:10.3847/0004-637X/818/2/132
  [arXiv:1511.00055 [astro-ph.CO]]
  %%CITATION = doi:10.3847/0004-637X/818/2/132;%%
  %16 citations counted in INSPIRE as of 10 Jul 2016 

%\cite{Bernal:2016gxb}
\bibitem{Bernal:2016gxb} 
  J.~L.~Bernal, L.~Verde and A.~G.~Riess,
  %``The trouble with $H_0$,''
  JCAP {\bf 1610}, no. 10, 019 (2016)
 % doi:10.1088/1475-7516/2016/10/019
  [arXiv:1607.05617 [astro-ph.CO]].
  %%CITATION = doi:10.1088/1475-7516/2016/10/019;%%
  %17 citations counted in INSPIRE as of 07 Jan 2017



%\cite{Sola:2016jky}
\bibitem{Sola1} 
  J.~Sola, A.~Gomez-Valent and J.~de Cruz Perez,
  %``First evidence of running cosmic vacuum: challenging the concordance model,''
  Astrophys.\ J.\  {\bf 836}, no. 1, 43 (2017)
  doi:10.3847/1538-4357/836/1/43
  [arXiv:1602.02103 [astro-ph.CO]].
  %%CITATION = doi:10.3847/1538-4357/836/1/43;%%
  %18 citations counted in INSPIRE as of 06 Apr 2017
  
%\cite{Sola:2017jbl}
\bibitem{Sola2} 
  J.~Sola, J.~d.~C.~Perez and A.~Gomez-Valent,
  %``Towards the firsts compelling signs of vacuum dynamics in modern cosmological observations,''
  arXiv:1703.08218 [astro-ph.CO].
  %%CITATION = ARXIV:1703.08218;%%
  %3 citations counted in INSPIRE as of 06 Apr 2017

%\cite{DiValentino:2017zyq}
\bibitem{DMLS} 
  E.~Di Valentino, A.~Melchiorri, E.~V.~Linder and J.~Silk,
  %``Constraining Dark Energy Dynamics in Extended Parameter Space,''
  arXiv:1704.00762 [astro-ph.CO].
  %%CITATION = ARXIV:1704.00762;%%

%\cite{DiValentino:2016hlg}
\bibitem{DMS} 
  E.~Di Valentino, A.~Melchiorri and J.~Silk,
  %``Reconciling Planck with the local value of $H_0$ in extended parameter space,''
  Phys.\ Lett.\ B {\bf 761}, 242 (2016)
  doi:10.1016/j.physletb.2016.08.043
  [arXiv:1606.00634 [astro-ph.CO]].
  %%CITATION = doi:10.1016/j.physletb.2016.08.043;%%
  %36 citations counted in INSPIRE as of 06 Apr 2017  
  
\bibitem{Kullback:1951va}
	S.~Kullback and R.~A.~Leibler
	Ann.\ Math.\ Stat. {\bf 22} (1951)  
  

%\cite{Zunckel:2008ti}
\bibitem{Om1} 
  C.~Zunckel and C.~Clarkson,
  %``Consistency Tests for the Cosmological Constant,''
  Phys.\ Rev.\ Lett.\  {\bf 101}, 181301 (2008)
  doi:10.1103/PhysRevLett.101.181301
  [arXiv:0807.4304 [astro-ph]].
  %%CITATION = doi:10.1103/PhysRevLett.101.181301;%%
  %49 citations counted in INSPIRE as of 06 Apr 2017

%\cite{Sahni:2008xx}
\bibitem{Om2} 
  V.~Sahni, A.~Shafieloo and A.~A.~Starobinsky,
  %``Two new diagnostics of dark energy,''
  Phys.\ Rev.\ D {\bf 78}, 103502 (2008)
  doi:10.1103/PhysRevD.78.103502
  [arXiv:0807.3548 [astro-ph]].
  %%CITATION = doi:10.1103/PhysRevD.78.103502;%%
  %196 citations counted in INSPIRE as of 25 Oct 2016
  
 %\cite{Pantazis:2016nky}
\bibitem{PNP} 
  G.~Pantazis, S.~Nesseris and L.~Perivolaropoulos,
  %``Comparison of thawing and freezing dark energy parametrizations,''
  Phys.\ Rev.\ D {\bf 93}, no. 10, 103503 (2016)
  doi:10.1103/PhysRevD.93.103503
  [arXiv:1603.02164 [astro-ph.CO]].
  %%CITATION = doi:10.1103/PhysRevD.93.103503;%%
  %10 citations counted in INSPIRE as of 06 Apr 2017   
  
%\cite{Ade:2015rim}
\bibitem{Ade:2015rim} 
  P.~A.~R.~Ade {\it et al.} [Planck Collaboration],
  %``Planck 2015 results. XIV. Dark energy and modified gravity,''
  Astron.\ Astrophys.\  {\bf 594}, A14 (2016)
  doi:10.1051/0004-6361/201525814
  [arXiv:1502.01590 [astro-ph.CO]].
  %%CITATION = doi:10.1051/0004-6361/201525814;%%
  %246 citations counted in INSPIRE as of 23 Mar 2017
  
  
%\cite{Chevallier:2000qy}
\bibitem{CP} 
  M.~Chevallier and D.~Polarski,
  %``Accelerating universes with scaling dark matter,''
  Int.\ J.\ Mod.\ Phys.\ D {\bf 10}, 213 (2001)
  doi:10.1142/S0218271801000822
  [gr-qc/0009008].
  %%CITATION = doi:10.1142/S0218271801000822;%%
  %1055 citations counted in INSPIRE as of 14 Nov 2016

%\cite{Linder:2002et}
\bibitem{L} 
  E.~V.~Linder,
  %``Exploring the expansion history of the universe,''
  Phys.\ Rev.\ Lett.\  {\bf 90}, 091301 (2003)
  doi:10.1103/PhysRevLett.90.091301
  [astro-ph/0208512].
  %%CITATION = doi:10.1103/PhysRevLett.90.091301;%%
  %1012 citations counted in INSPIRE as of 21 Mar 2017  
  
  
\bibitem{Feng:2004ff} 
  B.~Feng, M.~Li, Y.~S.~Piao and X.~Zhang,
  %``Oscillating quintom and the recurrent universe,''
  Phys.\ Lett.\ B {\bf 634}, 101 (2006)
  doi:10.1016/j.physletb.2006.01.066
  [astro-ph/0407432].
  %%CITATION = doi:10.1016/j.physletb.2006.01.066;%%
  %311 citations counted in INSPIRE as of 14 Nov 2016  

%\cite{Zhao:2016das}
\bibitem{BAOZhao} 
  G.~B.~Zhao {\it et al.},
  %``The clustering of galaxies in the completed SDSS-III Baryon Oscillation Spectroscopic Survey: tomographic BAO analysis of DR12 combined sample in Fourier space,''
  arXiv:1607.03153 [astro-ph.CO].
  %%CITATION = ARXIV:1607.03153;%%
  
%\cite{Wang:2016wjr}
\bibitem{BAOWang} 
  Y.~Wang {\it et al.},
  %``The clustering of galaxies in the completed SDSS-III Baryon Oscillation Spectroscopic Survey: tomographic BAO analysis of DR12 combined sample in configuration space,''
  arXiv:1607.03154 [astro-ph.CO].
  %%CITATION = ARXIV:1607.03154;%%

%\cite{Delubac:2014aqe}
\bibitem{Delubac14} 
  T.~Delubac {\it et al.} [BOSS Collaboration],
  %``Baryon acoustic oscillations in the Ly? forest of BOSS DR11 quasars,''
  Astron.\ Astrophys.\  {\bf 574}, A59 (2015)
%  doi:10.1051/0004-6361/201423969
  [arXiv:1404.1801 [astro-ph.CO]].
  %%CITATION = doi:10.1051/0004-6361/201423969;%%
  %153 citations counted in INSPIRE as of 10 Jul 2016   
  

%\cite{Beutler:2011hx}
\bibitem{6df} 
  F.~Beutler {\it et al.},
  %``The 6dF Galaxy Survey: Baryon Acoustic Oscillations and the Local Hubble Constant,''
  Mon.\ Not.\ Roy.\ Astron.\ Soc.\  {\bf 416}, 3017 (2011)
  %doi:10.1111/j.1365-2966.2011.19250.x
  [arXiv:1106.3366 [astro-ph.CO]].
  %%CITATION = doi:10.1111/j.1365-2966.2011.19250.x;%%
  %595 citations counted in INSPIRE as of 30 Jul 2016

%\cite{Ross:2014qpa}
\bibitem{MGS} 
  A.~J.~Ross, L.~Samushia, C.~Howlett, W.~J.~Percival, A.~Burden and M.~Manera,
  %``The clustering of the SDSS DR7 main Galaxy sample ? I. A 4 per cent distance measure at $z = 0.15$,''
  Mon.\ Not.\ Roy.\ Astron.\ Soc.\  {\bf 449}, no. 1, 835 (2015)
  %doi:10.1093/mnras/stv154
  [arXiv:1409.3242 [astro-ph.CO]].
  %%CITATION = doi:10.1093/mnras/stv154;%%
  %122 citations counted in INSPIRE as of 30 Jul 2016
  
  
%\cite{Parkinson:2012vd}
\bibitem{wigglez} 
  D.~Parkinson {\it et al.},
  %``The WiggleZ Dark Energy Survey: Final data release and cosmological results,''
  Phys.\ Rev.\ D {\bf 86}, 103518 (2012)
  doi:10.1103/PhysRevD.86.103518
  [arXiv:1210.2130 [astro-ph.CO]].
  %%CITATION = doi:10.1103/PhysRevD.86.103518;%%
  %102 citations counted in INSPIRE as of 29 Jun 2016

  %\cite{Riess:2016jrr}
\bibitem{R16} 
  A.~G.~Riess {\it et al.},
  %``A 2.4% Determination of the Local Value of the Hubble Constant,''
  arXiv:1604.01424 [astro-ph.CO].
  %%CITATION = ARXIV:1604.01424;%%
  %34 citations counted in INSPIRE as of 29 Jun 2016

%\cite{Moresco:2016mzx}
\bibitem{hz16} 
  M.~Moresco {\it et al.},
  %``A 6% measurement of the Hubble parameter at $z\sim0.45$: direct evidence of the epoch of cosmic re-acceleration,''
  JCAP {\bf 1605}, no. 05, 014 (2016)
  doi:10.1088/1475-7516/2016/05/014
  [arXiv:1601.01701 [astro-ph.CO]].
  %%CITATION = doi:10.1088/1475-7516/2016/05/014;%%
  %14 citations counted in INSPIRE as of 29 Jun 2016
  
  %\cite{Betoule:2014frx}
\bibitem{Betoule:2014frx} 
  M.~Betoule {\it et al.} [SDSS Collaboration],
  %``Improved cosmological constraints from a joint analysis of the SDSS-II and SNLS supernova samples,''
  Astron.\ Astrophys.\  {\bf 568}, A22 (2014)
 % doi:10.1051/0004-6361/201423413
  [arXiv:1401.4064 [astro-ph.CO]].
  %%CITATION = doi:10.1051/0004-6361/201423413;%%
  %265 citations counted in INSPIRE as of 10 Jul 2016  

%\cite{Heymans:2013fya}
\bibitem{WL} 
  C.~Heymans {\it et al.},
  %``CFHTLenS tomographic weak lensing cosmological parameter constraints: Mitigating the impact of intrinsic galaxy alignments,''
  Mon.\ Not.\ Roy.\ Astron.\ Soc.\  {\bf 432}, 2433 (2013)
%  doi:10.1093/mnras/stt601
  [arXiv:1303.1808 [astro-ph.CO]].
  %%CITATION = doi:10.1093/mnras/stt601;%%
  %162 citations counted in INSPIRE as of 10 Jul 2016
  
\bibitem{DESI} 
  A.~Aghamousa {\it et al.} [DESI Collaboration],
  %``The DESI Experiment Part I: Science,Targeting, and Survey Design,''
  arXiv:1611.00036 [astro-ph.IM].
  %%CITATION = ARXIV:1611.00036;%%
  %9 citations counted in INSPIRE as of 01 Jan 2017

%\cite{Astier:2010qf}
\bibitem{EuclidSN} 
  P.~Astier, J.~Guy, R.~Pain and C.~Balland,
  %``Dark energy constraints from a space-based supernova survey,''
  Astron.\ Astrophys.\  {\bf 525}, A7 (2011)
 % doi:10.1051/0004-6361/201015044.
  [arXiv:1010.0509 [astro-ph.CO]].
  %%CITATION = doi:10.1051/0004-6361/201015044;%%
  %13 citations counted in INSPIRE as of 01 Jan 2017    
  
\bibitem{CAMB} 
  A.~Lewis, A.~Challinor and A.~Lasenby,
  %``Efficient computation of CMB anisotropies in closed FRW models,''
  Astrophys.\ J.\  {\bf 538}, 473 (2000).  Available at \url{http://camb.info}
  [arXiv:astro-ph/9911177].
  %%CITATION = ASTRO-PH/9911177;%%
  
\bibitem{DEP} G.~B.~Zhao {\it et al.}, 
%J.~Q.~Xia, M.~Li, B.~Feng and X.~Zhang, 
Phys.\ Rev.\  D {\bf 72}, 123515 (2005) 
[arXiv:astro-ph/0507482]    
  
%\cite{Lewis:2002ah}
\bibitem{Lewis:2002ah}
  A.~Lewis and S.~Bridle,
  %``Cosmological parameters from CMB and other data: a Monte-Carlo approach,''
  Phys.\ Rev.\  D {\bf 66} (2002) 103511
  [arXiv:astro-ph/0205436].
  %%CITATION = PHRVA,D66,103511;%%    
  

%\cite{Handley:2015fda}
\bibitem{polychord} 
  W.~J.~Handley, M.~P.~Hobson and A.~N.~Lasenby,
  %``PolyChord: nested sampling for cosmology,''
  Mon.\ Not.\ Roy.\ Astron.\ Soc.\  {\bf 450}, no. 1, L61 (2015)
  doi:10.1093/mnrasl/slv047
  [arXiv:1502.01856 [astro-ph.CO]].
  %%CITATION = doi:10.1093/mnrasl/slv047;%%
  %11 citations counted in INSPIRE as of 29 Jun 2016    
  

\end{thebibliography}
\end{document}